# An Efficient Production Process for Extracting Salivary Glands from Mosquitoes

Mariah Schrum, Amanda Canezin, Sumana Chakravarty, Michelle Laskowski, Suat Comert, Yunuscan Sevimli, Gregory S. Chirikjian, Stephen L. Hoffman, and Russell H. Taylor*

*Abstract*— Malaria is the one of the leading causes of morbidity and mortality in many developing countries. The development of a highly effective and readily deployable vaccine represents a major goal for world health. There has been recent progress in developing a clinically effective vaccine manufactured using *Plasmodium falciparum* sporozoites (PfSPZ) extracted from the salivary glands of Anopheles sp. Mosquitoes. The harvesting of PfSPZ requires dissection of the mosquito and manual removal of the salivary glands from each mosquito by trained technicians. While PfSPZ-based vaccines have shown highly promising results, the process of dissection of salivary glands is tedious and labor intensive. We propose a mechanical device that will greatly increase the rate of mosquito dissection and deskill the process to make malaria vaccines more affordable and more readily available. This device consists of several components: a sorting stage in which the mosquitoes are sorted into slots, a cutting stage in which the heads are removed, and a squeezing stage in which the salivary glands are extracted and collected. This method allows mosquitoes to be dissected twenty at a time instead of one by one as previously done and significantly reduces the dissection time per mosquito.

*Index Terms*— Biomedical device, design automation, malaria, vaccine, production fixturing, productivity enhancement, production learning curve.

| Author | Email | Phone Number |
|---|---|---|
| Mariah Schrum | mschrum2@jhu.edu | 412-327-9957 |
| Amanda Canezin | acanezi2@jhu.edu | 516-509-9969 |
| Sumana Chakravarty | schakravarty@sanaria.com | 240-403-2723 |
| Michelle Laskowski | mlaskowski@sanaria.com | 208-908-8792 |
| Suat Comert | scomert1@jhu.edu | 443-447-6664 |
| Yunuscan Sevimli | yunus@jhu.edu | 443-835-6628 |
| Gregory S. Chirikjian | gchirik1@jhu.edu | 410- 516-7127 |
| Stephen L. Hoffman | slhoffman@sanaria.com | 240-403-2701 |
| Russell H. Taylor | rht@jhu.edu | 410-516-6299 |

*Acknowledgements:* The work reported was supported in part by a research subcontract to Johns Hopkins University from Sanaria, Inc. under NIH SBIR Grant R43AI112165-01, and in part by Johns Hopkins University and Sanaria internal funds.

*Affiliations:* M. Schrum, G. Chirikjian, and R. Taylor are affiliated with the Laboratory for Computational Sensing and Robotics (LCSR) at Johns Hopkins University. Previously, A. Canezin, S. Coemert and Y. Sevimli were also with LCSR. A. Canezin is now with Accenture, S. Coemert is with T.U. Munich and Y. Sevimli is with Galen Robotics, Inc.
S, Chakravarty and S.L. Hoffman are with Sanaria, Inc. Rockville, Md. At the time the reported work was done M. Laskowski was also with Sanaria. She is now with Leidos Life Sciences, Frederick, Maryland, USA.

* Corresponding Author

## 1. INTRODUCTION

Malaria remains one of the most important infectious diseases in the world. The World Health Organization estimates that there were 216 million cases of malaria worldwide in 2016, with 438,000 deaths [2, 3]. *Plasmodium falciparum* (Pf) is responsible for more than 98% of all deaths from malaria. Development of vaccines against all of the Plasmodium species parasites that cause malaria is a significant public health priority. However, the major priority is developing a vaccine against Pf.

Humans become infected with malaria-causing parasites when Anopheles mosquitoes inoculate the "sporozoite (SPZ)" developmental stage of the parasite. SPZ resides in mosquito salivary glands immediately prior to passage to humans during feeding (Fig 1.1). Currently, there is no licensed vaccine for the prevention of malaria. However, significant progress has been made in developing whole PfSPZ vaccines [4-8]. These vaccines have shown a high level protective efficacy against controlled human malaria infection and malaria transmitted in the field. Their safety and high-level efficacy will make them ideal for large-scale malaria elimination campaigns in geographically defined malarious regions. However, a limiting step in the manufacture of PfSPZ-based vaccines has been the extraction of the salivary glands and isolation of sporozoites from very large numbers of infected mosquitoes to meet expected demand for the vaccine.

The current gland extraction method is manual, time consuming, and labor intensive. To gain access to the glands, a technician first removes the mosquito head using the beveled edge of a hypodermic needle as a knife. Next, the technician gently squeezes the mosquito body to remove the gland from the thorax. Once a pool of 30-50 salivary glands are extracted, an operator suctions them using a pipettor and transfers them to a collection tube. This method produces a throughput of around 5-6 mosquitoes per minute on average. It requires extensive training and practice over several months to achieve this rate.

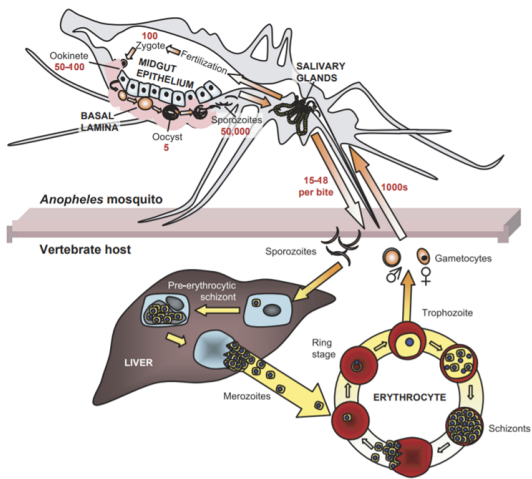

**Fig 1. Life cycle of *plasmodium falciparum*.** Note the location of the salivary glands in the mosquito. [1]

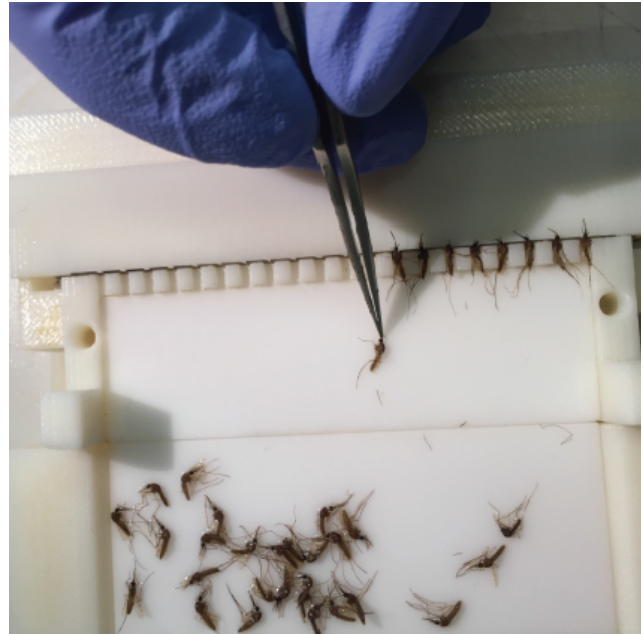

**Fig. 2. Sorting mosquitoes into cartridge slots.** The technician grasps each mosquito by its proboscis and drags it into a slot.

There has been at least one effort to fully automate the salivary gland extraction production process by means of a robot [9-11], but the robot was never completed. Although there is very little technical material publicly available about this system beyond a YouTube video included as part of a fund-raising effort [10] and a newspaper article [11], the proposed system evidently included a computer vision system to locate mosquitoes in a dish and a Cartesian (XYZ) robot to position an end effector that would grasp the mosquito and feed each one serially and sequentially into a tube for further processing.

Although the ultimate goal is a fully automated mosquito dissection and gland extraction process, the immediate goal is more limited: to significantly increase the productivity of manual dissection by human technicians while providing experience for future automation. To accomplish this, we have developed a simple, modular fixturing system that allows several time-consuming steps of the process to be performed concurrently on multiple mosquitoes, while also greatly simplifying the remaining per-mosquito actions performed by the human technician. Equally important, it dramatically reduces the time required to train an operator to perform the procedure.

## 2. MATERIALS AND METHODS

In developing our semi-automation approach, we recognized that the fundamental problem was to align each mosquito so that the decapitation and gland extraction steps could be automated without needless complications from extraneous mosquito parts such as legs and wings. Further, this could be accomplished with relatively simple fixturing enabling the technician to load batches of mosquitoes into cartridges aligning them so that subsequent steps could be performed in parallel.

Our current fixture design (shown in Fig. ) consists of several modular components, including sorting cartridges, blade assembly, squeezer, and staging area. Each sorting cartridge has 20 slots allowing for the dissection of 20 mosquitoes at a time. A slot is 1.25 mm wide, making it slightly wider than a mosquito body so that the mosquito can easily be placed in the slot but still held in position during the subsequent stages. The slot length and depth are 3mm and 1.5mm respectively. It was found that slots with dimensions slightly larger than the size of the mosquito were most effective for keeping the mosquito well aligned during the squeezing stage of the dissection. Because the surface over which the mosquitoes are dragged must be smooth, each cartridge has a staging area made of acrylic. The staging area is 71mm by 22mm. About 30 to 40 mosquitoes can be spread out over this area which is a sufficient number to efficiently fill the cartridges. The sorting cartridge is removable so that when the salivary glands are extracted, they do not become trapped behind the blades.

The blades have notches approximately 0.6 mm wide in which the mosquito neck sits. Like clippers, the blades slide past one another, cutting off each mosquito head simultaneously. Because the mosquito's neck is contacted by a sharp edge on both sides, this blade design ensures a clean cut. The blades are spring loaded so that they sit flush against the sorter cartridge. Both blades are removable, allowing for them to be easily cleaned or simply replaced. The blades are made of 50μm thick stainless steel and are thin enough that sharpening is not necessary.

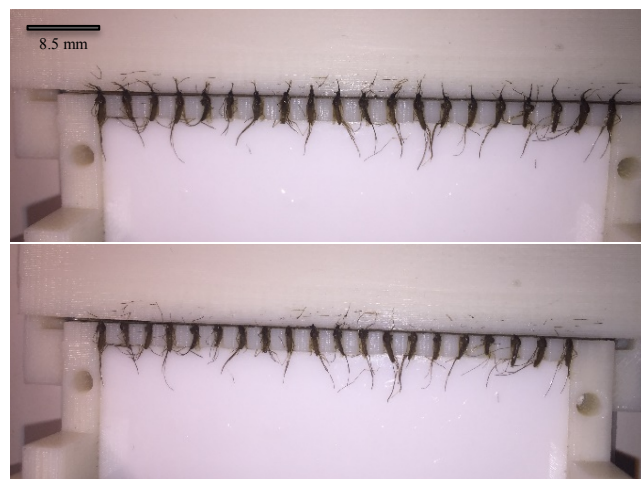

**Fig. 4. Decapitation**. (Top) Mosquitoes aligned in cartridge with heads between decapitation blades. (Bottom) Mosquitoes after decapitation.

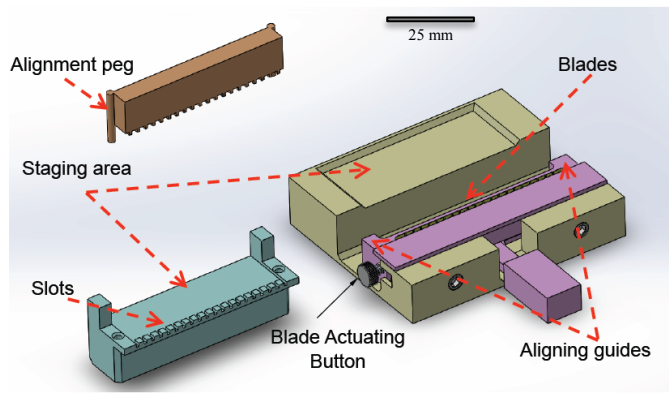

**Fig. 3. Design of mosquito gland extraction apparatus,** including the sorting cartridge (blue), the blade assembly (pink), squeezer (brown) and staging area (tan).

The squeezer is comb-like, with 20 rectangular teeth that fit into each of the slots on the sorter. To ensure the glands are extracted, the squeezer must contact each mosquito in the thorax region. Therefore, the squeezer has two round pegs that fit into matching holes in the sorter to guarantee perfect alignment. The teeth fit in the slots with very little clearance to ensure that the glands are extruded forward and do not become trapped between the side of the slot and the squeezer tooth.

The workflow proceeds as follows: First, a cartridge is inserted into the apparatus so that the slots align with the blade openings, and a clump of mosquitoes is placed onto the staging area. A small amount of an aqueous medium is also placed onto the staging area. Using tweezers, the technician grasps mosquitoes one-at-a-time by the proboscis and drags the mosquito into a cartridge slot and places it so that the mosquito neck is between the clipper blades, as shown in Fig. 2. This process causes the legs and wings of the mosquito to fold back along the mosquito body, where they are constrained by the cartridge slot. This process is repeated until all cartridge slots are filled.

Once the necks are aligned between the blades, the blades are actuated manually via a button on the left side of the device (labeled in Fig. ), enabling them to slide past one another and sever the neck of each mosquito. Fig. 4 depicts the mosquitoes before and after the heads have been removed. Next, the sorting cartridge is removed and another empty sorting cartridge can be inserted.

Finally, the glands are extracted. To do this, the squeezer comb is aligned with the cartridge by placing its aligning posts into the corresponding holes in the cartridge so that the teeth on the comb rest on the mosquito thoraces. The technician then presses down to squeeze the glands out of the thoraces. The glands are ejected from the mosquito onto the flat surface in front of the blades where they can be collected by a pipettor and placed into collection tubes.

The apparatus described in this section will be referred to as a semi-automated mosquito microdissection system (SAMMS). Our goal in the long run is to create a fully automated system, with minimal to no manual input.

**Table 1: Training time required for previous (unassisted) process**

| Operator | 1 | 2 | 3 | 4 | 5 | 6 | 7 | 8 | 9 | 10 | 11 | 12 | 13 | 14 | 15 | 16 | 17 | 18 | 19 | 20 | 21 | 22 | 23 | 24 | 25 | **Avg** |
|---|---|---|---|---|---|---|---|---|---|---|---|---|---|---|---|---|---|---|---|---|---|---|---|---|---|---|
| Weeks training | 48 | 9 | 27 | 30 | 55 | 7 | 30 | 17 | 32 | 19 | 9 | 10 | 3 | 50 | 48 | 37 | 31 | 9 | 11 | 24 | 141 | 15 | 24 | 36 | 14 | **29** |

## 3. EXPERIMENTS AND RESULTS

*Quality of salivary gland SPZ.* The semi-automated mosquito micro-dissection system recapitulates the manual dissection procedure in its entirety, such that the forces exerted on mosquitoes when using it are nearly identical to those already in effect in the fully manual process. Prior to late 2015 the only digression from the original manual protocol in place was the use of pipettes to collect salivary glands by aspiration into designated tips, and subsequent discharge into 1.5 mL collection tubes. The manual method at the time involved a pick-and place approach for the extruded glands gathering 3-5 pairs of glands on the tip of a needle and placing them in fluid contained in collection tubes. However, as part of a change in workflow and configuration of manual dissection, we successfully transitioned to a method that applies the suction approach to collecting salivary glands, identical to the method being used with the semi-automated prototype described here. Fig. 5 shows comparative results of mosquito processing capacity using the old and new methods in 2015 alone.

*Semi-automated workflow radically reduces training period to qualify operators for mosquito microdissection.* In order to minimally dissect 100 mosquitoes per hour and be certified for manual dissection at Sanaria, untrained personnel go through a rigorous training procedure involving 1-3 one-hour sessions every week. The time to complete this training has varied tremendously between operators depending on dexterity, and hand-eye coordination skills for successful micromanipulation of mosquitoes under a stereomicroscope (Table 1). Although further entrainment does occur for every operator as they continue participation in dissection, the gestation period prior to qualification is too variable and long and averages around 29 weeks.

In contrast, although the rate-limiting step in operation of SAMMS for un-trained operators was the time to load mosquitoes into decapitation cartridges, even in their very first trials, operator times ranged from 338-649 mosquitoes per hour. Entrainment occurred over as little as 3 trials (Table 2). As trainees practiced over successive days, every trainee achieved a total mosquito alignment and decapitation time for a 20 mosquito cartridge of less than 1.5 minutes and gland collection time of 0.5-1.0 minute, similar to operator 6, (Table 2) and a total output of 600 mosquitoes per hour. The number of iterations/training sessions for individuals to achieve this target was fairly uniform taking one day (3–10 trials over one hour) to 3 days for completion of training. This was a very significant improvement over typical training periods for manual dissection described above and entails a 10-15 fold reduction in training time.

*Targeting 600-mosquito processing capacity per hour allows at least a two fold increase in throughput over current manual dissection.* The average rate of dissecting mosquitoes by completely manual methods averages around 320 mosquitoes with a wide range in individual operator capabilities (260 to 420 mosquitoes) as shown in Fig 5 (new method 2015 and 2016). The ability to quickly master a 600-mosquito processing capacity with the semi-automated device as described above represents at least a two-fold increase in throughput over current rates.

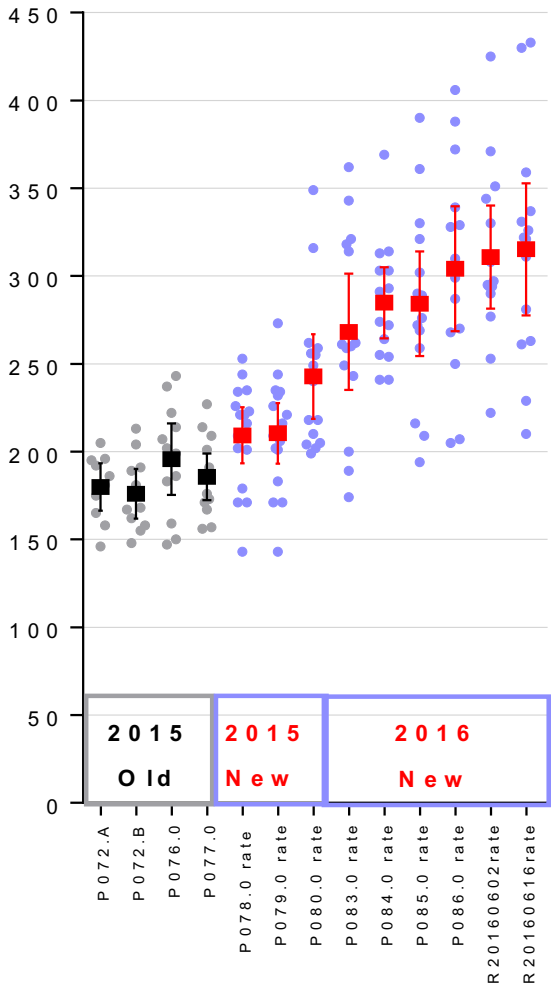

**Fig. 5: Rate of dissection of mosquito salivary glands by manual dissection.** The rate is calculated as the number of Mosquitoes Dissected Per Hour (Mdph). Individual grey or blue scatter points represent a single dissector's average recorded on one production day. The mean (black or red bars) of all participating dissectors in a specific production day is plotted with 95% confidence intervals. In an earlier iteration all steps in dissection used to be performed by trained dissectors (grey scatter points, black bars). In a new modified configuration (blue scatter points, red bars) only the task of decapitation and separation of glands from mosquito bodies was conducted by highly trained dissectors and glands were pooled and collected by separate less-specialized personnel.

**Table 2: Operation times and production rates for 8 operators using a proposed apparatus. (A)** Minutes to align 20 mosquitoes; **(B)** Minutes for gland extrusion and collection. Total time for 20 mosquitoes is in minutes, and rate is mosquito throughput per hour.

|       | 1   | 2   | 3   | 4   | 5   | 6   | 7   | 8   | Avg  |
|-------|-----|-----|-----|-----|-----|-----|-----|-----|------|
| (A)   | 2.4 | 2.5 | 2.3 | 2.5 | 1.5 | 1.3 | 1.1 | 1.2 | **1.9** |
| (B)   | 0.7 | 0.8 | 0.5 | 1.1 | 1.2 | 0.7 | 1.1 | 0.7 | **0.8** |
| Total | 3.1 | 3.3 | 2.8 | 3.6 | 2.7 | 2   | 2.2 | 1.9 | **2.7** |
| Rate  | 393 | 364 | 429 | 338 | 444 | 600 | 545 | 649 | **470** |

## 4. DISCUSSION

As mentioned in the Introduction, the immediate goal was to develop production fixtures and methods that could significantly improve the existing manual process while providing the experience necessary to develop a fully automated mosquito dissection and gland extraction process. Accordingly, we have developed prototypes for robust mosquito alignment, decapitation and gland extraction, and integrated into a semi-automated mosquito micro-dissection system.

The current decapitation system allows batch-processing of 20 to potentially hundreds of mosquitoes at a time. Despite this range, we believe optimum use during an actual production campaign at Sanaria would be limited to processing a maximum of about 40 mosquitoes per cartridge due to specimen drying issues and the balance between ease of use and throughput. Gland extraction from mosquito thoraces is also vastly simplified, again allowing batch-processing from tens to hundreds of mosquitoes at once. As a result, mosquito processing capacity by one-operator using one unit of the semi-automated prototype compared to a single operator performing manual dissection is expected to double, or even triple over future operator entrainment cycles. Even further increase in efficiency is expected through leveraging alternative task-distribution methods that have already been implemented for manual dissection and yielded at least a 2.5 fold increase in throughput between late 2015 and into 2016. The single most useful advantage of the development of the SAMMS is that training time to qualify individual operators for mosquito processing is radically reduced by 15-20 fold, compared to manual dissection. This, together with reduction of fatigue in individuals involved with semi-automated processing is expected to eliminate any hypothetical drop-out rates in the task-force ascribed to this specific task. Furthermore, the yield of sporozoites recovered per mosquito using SAMMS has been comparable, if not better than, that achieved with manual dissection.

Developing our current apparatus required considerable experimentation with alternative designs before we converged the embodiment described herein, and we will discuss these experiences briefly. The approach of grasping each mosquito's proboscis and dragging it across a surface lubricated with aqueous culture medium into an appropriate cartridge slot was very successful from the beginning. We explored a number of different decapitation methods and apparatus before settling on the clipper blades arrangement described above. This method reliably severs the necks with minimum displacement of the mosquito bodies. Also, our human technicians do not find it difficult to guide mosquitoes so that the necks are properly positioned between the blades. As discussed below, these experiences have led us to explore a vision-driven robotic process for these aspects of our full automation approach. Similarly, we explored several alternative approaches for salivary gland extraction and collection before finally settling on our current comb squeezer and pipette collection approach. In particular, we found that having a fairly tight fit between the squeezer comb teeth and the cartridge slots ensures that the gland material is extruded to the front surface of the cartridge. At this point, the glands tend to stick to the front surface, and the suction device can easily gather them, as well as any that have slid down to the bottom of the cartridge.

## 5. CONCLUSIONS AND FUTURE WORK

In this paper, we have presented the design and workflow for a semi-automated mosquito dissection system that addresses a significant problem for the production of a malaria vaccine. Our prototype design is currently being refined with regards to ease of use in a multi-operator user interface, for smoother operations under continuous workflow spanning 8-10 hours, and materials compliance for manufacturing under current good manufacturing practices (cGMPs) as specified by the FDA. We are in the process of implementing a cGMP-compliant version of SAMMS to extract mosquito salivary glands in our production scheme for phase III clinical trials. The cost of the total number of SAMMS units required to process 50,000 mosquitoes, in order to produce one lot of vaccine is estimated to be ~$1500.

Aside from its near-term value in increasing production productivity while also significantly reducing operator fatigue and training time, our semi-automated approach represents an important stepping-stone toward development of a fully automated system, incorporating many of the lessons that we have learned in this initial project. We are beginning to develop such a system, which will rely on computer vision, small pick-and-place robotic devices, and an advanced feeding system for mosquito bodies.